\documentclass[aps,pre,superscriptaddress,twocolumn,balancelastpage,
               longbibliography]{revtex4-2}

\usepackage{times}
\usepackage{amsmath,amssymb}
\usepackage{graphicx}
\graphicspath{{Figures/}{/}}

\usepackage[english]{babel}
\usepackage{verbatim}
\usepackage{color}

\usepackage{placeins}    
\usepackage{flafter}     

\usepackage[space]{grffile}

\usepackage{color}
\definecolor{darkblue}{rgb}{0., 0., 0.55}
\usepackage[colorlinks=true, urlcolor=darkblue, citecolor=darkblue, linkcolor=darkblue]{hyperref}

\usepackage[normalem]{ulem}
\definecolor{dartmouthgreen}{rgb}{0.05, 0.5, 0.06}

\newcommand{\ue}{\text{e}}          
\newcommand{\ui}{\text{i}}          
\newcommand{\ud}{\text{d}}          

\newcommand{\xbold}{\boldsymbol{x}}          
\newcommand{\bbold}{\boldsymbol{b}}          
\newcommand{\reflbold}{\boldsymbol{r}}          

\newcommand{\ginv}{\ensuremath{\gamma_{\text{inv}}}}
\newcommand{\gnat}{\ensuremath{\gamma_{\text{nat}}}}
\newcommand{\gtyp}{\ensuremath{\gamma_{\text{typ}}}}

\newcommand{\rhoinv}{\ensuremath{\rho_{\text{inv}}}}
\newcommand{\rhonat}{\ensuremath{\rho_{\text{nat}}}}
\newcommand{\rhoxi}{\ensuremath{\rho_{\gamma}}}

\newcommand{\hus}[1]{\ensuremath{H}_{#1}}       

\newcommand{\ev}[1]{\langle #1 \rangle}         
\newcommand{\tHpsi}{\widetilde{H}_{\psi}}       
\newcommand{\aHg}{\ev{\hus{}}_\gamma}
\newcommand{\navg}{n_\text{avg}}
\newcommand{\nsample}{n_\text{s}}

\newcommand{\meanI}[1][\varphi]{\mu(#1, \gamma)}
\newcommand{\Iphi}[1][\varphi]{\ensuremath{I_{#1}(\psi)}}
\newcommand{\tilIphi}[1][\varphi]{\ensuremath{\tilde{I}_#1(\psi)}}

\newcommand{\densMax}{I_\text{c}}

\newcommand{\op}{\ensuremath{\text{Op}_N}}              
\newcommand{\braket}[2]{\langle #1 | #2 \rangle}        

\newcommand{\refl}{\ensuremath{r}}
\newcommand{\reflOmega}{\ensuremath{r_\Omega}}
\newcommand{\Qrefl}{\hat{\ensuremath{R}}}
\newcommand{\QMapcls}{\hat{\ensuremath{U}}}     
\newcommand{\MMapcls}{{\ensuremath{\hat{U}^{\text{cue}}}}}
\newcommand{\QMap}{\QMapcls_{\refl}}            
\newcommand{\MMap}{{\MMapcls_{\refl}}}
\newcommand{\Mapcls}{\ensuremath{M}}            
\newcommand{\BakerMap}{\ensuremath{B_{\bbold}}}            
\newcommand{\QBakerMap}{\ensuremath{\hat{B}_{\bbold}}}            

\newcommand{\PS}{\ensuremath{\Gamma}}           

\newcommand{\prob}[2][]{\operatorname{P}_{#1}\left(#2\right)}
\renewcommand{\prob}[2][]{P_{#1}\left(#2\right)}

\makeatletter
\let\Hy@backout\@gobble
\makeatother

\begin{document}

    \title{Universal intensity statistics of multifractal resonance states}
    
    \author{Konstantin Clau{\ss}}
    \affiliation{Technische Universit\"at Dresden,
        Institut f\"ur Theoretische Physik and Center for Dynamics,
        01062 Dresden, Germany}
    
    \author{Felix Kunzmann}
    \affiliation{Technische Universit\"at Dresden,
        Institut f\"ur Theoretische Physik and Center for Dynamics,
        01062 Dresden, Germany}
    
    \author{Arnd B\"acker}
    \affiliation{Technische Universit\"at Dresden,
        Institut f\"ur Theoretische Physik and Center for Dynamics,
        01062 Dresden, Germany}
    \affiliation{Max-Planck-Institut f\"ur Physik komplexer Systeme,
        N\"othnitzer Stra\ss e 38, 01187 Dresden, Germany}
    
    \author{Roland Ketzmerick}
    \affiliation{Technische Universit\"at Dresden,
        Institut f\"ur Theoretische Physik and Center for Dynamics,
        01062 Dresden, Germany}
    \affiliation{Max-Planck-Institut f\"ur Physik komplexer Systeme,
        N\"othnitzer Stra\ss e 38, 01187 Dresden, Germany}

    \date{\today}

    \begin{abstract}
        We conjecture that in chaotic quantum systems with
        escape
        the intensity statistics for resonance states universally
        follows an exponential distribution.
        This requires a scaling by the multifractal mean intensity
        which depends on the system
        and the decay rate of the resonance state.
        We numerically support the conjecture by studying
        the phase-space Husimi function and the position representation
        of resonance states of the chaotic standard map,
        the baker map, and a random matrix model, each with partial escape.

    \end{abstract}
    
    \maketitle

    \section{Introduction}
    
    A detailed understanding of the structure and
    fluctuations of eigenstates
    is essential for the description of complex systems.
    For closed systems with classically ergodic dynamics
    almost all quantum eigenstates converge weakly towards the uniform measure
    on phase space as proven by the quantum ergodicity theorem
    \cite{Shn1974,CdV1985,Zel1987,ZelZwo1996,BaeSchSti1998}.
    This uniform limit is also established
    for quantum maps \cite{DegGraIso1995,BouDeB1996}.
    More detailed information is provided by
    the statistical fluctuations of eigenstates.
    For quantum billiards the random wave model \cite{Ber1977b}
    implies a Gaussian distribution of the eigenstate amplitudes, leading to
    a universal
    exponential distribution of the intensities, as confirmed,
    e.g., in Refs.~\cite{McDKau1988, AurSte1991, LiRob1994,Pro1997b}.
    For quantum maps with fully chaotic classical dynamics
    the eigenvector statistics is expected to be described by those of random matrices,
    originally introduced to describe the
    statistics of transition strengths of complex nuclei
    \cite{BroFloFreMelPanWon1981,GuhMueWei1998}.
    For systems without symmetry this leads to an exponential distribution
    of the intensities, as demonstrated,
    e.g., in Refs.~\cite{Izr1987, ZycLen1991, KusMosHaa1988, NonVor1998, Bae2003}.
    Restricted random wave models have been proposed
    to describe for example systems with a mixed phase space
    \cite{BaeSch2002a,BaeNon:p} and
    other non-isotropic cases \cite{BieLepHel2003,UrbRic2003, UrbRic2006}.
    The statistical properties of eigenstates play
    an important role in the context of many-body systems,
    see Refs.~\cite{LucSca2013,BeuBaeMoeHaq2018,BaeHaqKha2019}
    and references therein.

    In general, physical systems are not completely closed.
    They often show (partial) loss of particles or intensity
    in some interaction region \cite{AltPorTel2013},
    e.g., as in the three-disk scattering system
    \cite{GasRic1989c,Wir1999,WeiBarKuhPolSch2014} or in optical microcavities \cite{CaoWie2015}.
    Such scattering systems are described by resonance poles and the
    corresponding resonance states $\psi$ which have a decay rate $\gamma$.
    The distribution of decay rates is given
    by a fractal Weyl law in systems with full escape
    \cite{Sjo1990,Lin2002,
        LuSriZwo2003,SchTwo2004,RamPraBorFar2009,
        EbeMaiWun2010,ErmShe2010,
        PedWisCarNov2012,NonSjoZwo2014},
    and has been studied in systems with partial escape
    \cite{WieMai2008,NonSch2008,GutOsi2015,SchAlt2015}.

    Resonance states of chaotic systems with escape are generally not
    uniformly distributed, e.g., see Fig.~\ref{FIG:ScaledHusimis}~(b).
    Instead, they show a strong dependence on
    the phase-space region and decay rate $\gamma$.
    The average structure of resonance states with
    decay rate $\gamma$ is described
    by a multifractal measure on phase space
    \cite{CasMasShe1999b,
        LeeRimRyuKwoChoKim2004,
        KeaNovPraSie2006,
        NonRub2007,
        KeaNonNovSie2008,         
        HarShi2015,
        KoeBaeKet2015,
        KulWie2016,
        ClaKoeBaeKet2018,
        ClaAltBaeKet2019,
        BitKimZenWanCao2020}.
    Such measures are 
    conditionally invariant \cite{DemYou2006,AltPorTel2013},
    i.e., invariant under the corresponding classical
    dynamics with escape up to a
    global decay with rate $\gamma$.
    The most recent classical construction of such measures
    describes the average structure
    of resonance states quite well, but still shows deviations in the semiclassical limit for
    most $\gamma$ \cite{ClaAltBaeKet2019}.

    Individual resonance states of systems with escape
    have been discussed in terms of scarring on periodic orbits
    \cite{Hel1984},
    for microcavities
    \cite{LeeRimRyuKwoChoKim2004,FanYamCao2005,Wie2006,WieHen2008}
    and quantum maps
    \cite{WisCar2008,ErmCarSar2009,
        NovPedWisCarKea2009,CarBenBor2016}.
    However, a systematic study of the statistical properties of individual resonance states is 
    still missing, even in fully chaotic systems with escape.
    In particular, the question arises if there are universal
    properties of the intensity statistics.

    These intensity fluctuations 
    occur in the position representation of resonance states in chaotic scattering systems like the three-disk billiard.
    They also play an important role in the lasing
    properties of optical microcavities, which depend on individual
    resonance states.
    Knowledge of the intensity statistics might also
    allow to distinguish enhancement due to fluctuations from enhancement due to scarring
    on periodic orbits.

    In this paper we conjecture that in chaotic quantum systems with
    escape 
    a suitably scaled intensity statistics for resonance states universally
    follows an exponential distribution with mean one.
    To this end we scale by the mean intensity
    which depends on the system, the decay rate of the resonance state,
    and the phase-space region.
    We numerically support the conjecture by studying
    the phase-space Husimi function
    and the position representation of resonance states
    of the standard map with chaotic dynamics,
    the baker map,
    and a random matrix model, each with partial escape.

    The paper is organized as follows.
    In Sec.~\ref{SEC:ScaledIntensities} we introduce the class of
    quantum maps with escape
    and illustrate the scaled intensities for resonance states.
    In Sec.~\ref{SEC:Conjecture} we propose a conjecture about
    the statistics of these scaled intensities.
    In Sec.~\ref{SEC:NumericalResults} we present numerical support
    for the conjecture in two exemplary quantum maps with escape
    and for a random matrix model.
    The results are summarized in Sec.~\ref{SEC:Conclusion}.
    

    \section{Scaled intensities of resonance states}
    \label{SEC:ScaledIntensities}
    We consider dynamical systems which are described by a
    time-discrete map $\Mapcls$ on a bounded phase space $\PS$,
    originating, e.g., from time-periodically driven systems or
    a Poincar{\'e} section of an autonomous system.
    Escape (or gain) is introduced for such a system
    by a reflectivity function $\refl\colon \PS \rightarrow \mathbb{R}_{\geq0}$,
    such that
    $\refl(\xbold)$ describes the factor by which the intensity
    at the phase-space point $\xbold$ changes per time step
    \cite{AltPorTel2013}.
    Regions with escape (gain) are described by $\refl < 1$ ($\refl > 1$).

    \begin{figure}[b!]
        \includegraphics[scale=1.]{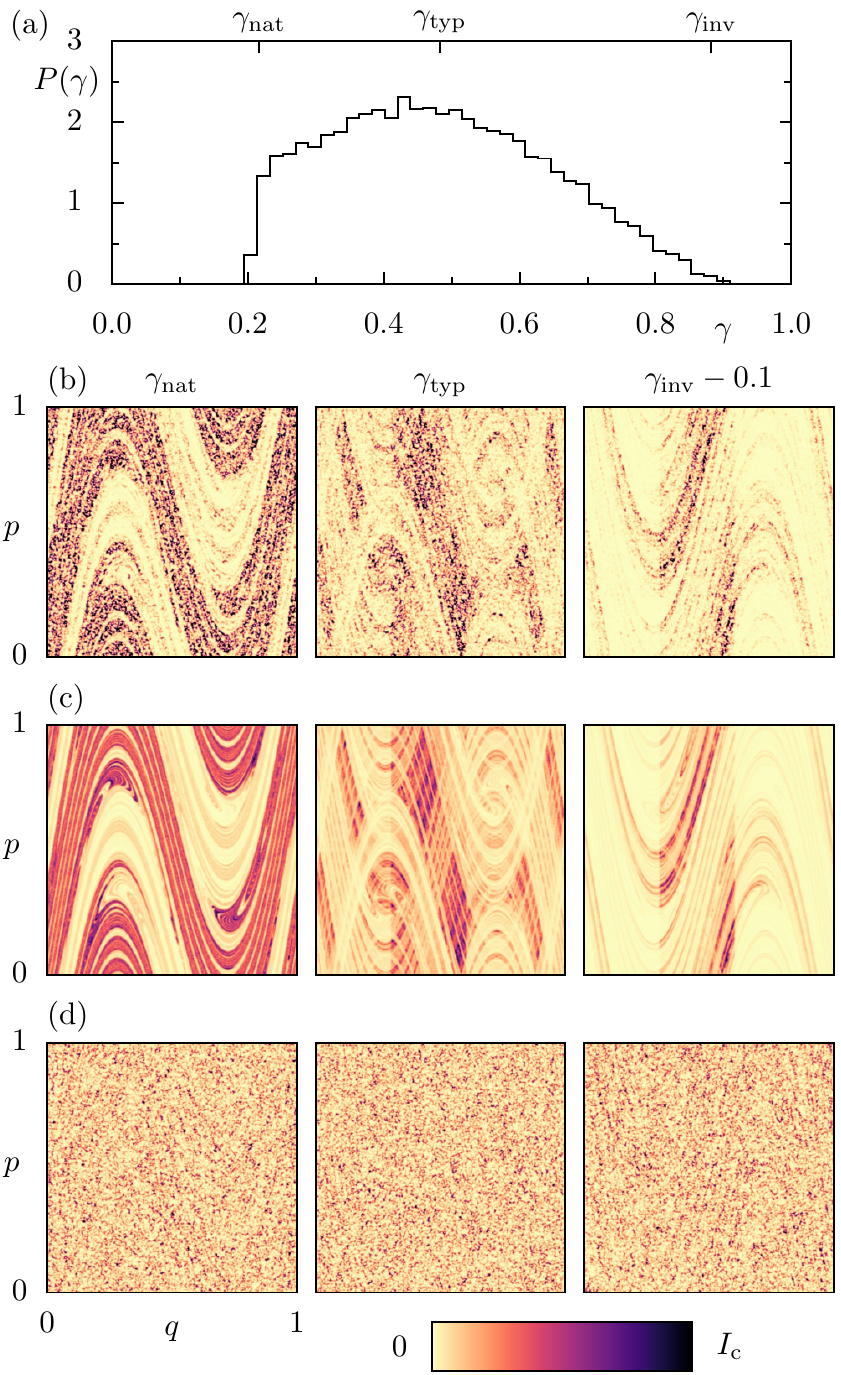}
        \caption{
            (a) Distribution of quantum decay rates $\gamma$
            for the chaotic standard map with partial escape,
            defined in App.~\ref{SEC:StandardMap}, with $h=1/16000$.
            Classical decay rates
            $\gnat\approx0.22, \gtyp\approx0.48, \ginv\approx0.88$ are
            indicated.
            (b)
            Husimi functions $\hus{\psi}$
            for three exemplary resonance states $\psi$ with decay rates
            $\gnat$, $\gtyp$, and $\ginv - 0.1$.
            (c) Averaged Husimi function $\aHg$ of
            $\navg = 200$ resonance states close to each $\gamma$.
            The same colormap with maximum $\densMax$
            is used for each pair of individual and
            average Husimi function, where
            $\densMax = \max_\PS \aHg$.
            (d) Scaled Husimi functions $\tHpsi$,
            see Eq.~\eqref{EQ:ScaledHusimi},
            visualized with $\densMax = 8$.
        }
        \label{FIG:ScaledHusimis}
    \end{figure}
    The corresponding quantum map with escape (or gain) is composed of
    the closed systems time evolution operator $\QMapcls$
    of dimension $N = 1/h$
    (quantizing the map $\Mapcls$ using an effective Planck's constant $h$)
    and some reflection operator $\Qrefl = \op{\sqrt{r}}$ (quantizing
    the reflectivity function $\refl$) \cite{DegGra2003b}.
    Without loss of generality we assume that escape takes place before
    the closed time evolution such that the quantum map with escape is defined as
    \begin{equation}
    \QMap = \QMapcls\Qrefl .
    \label{EQ:QMapEscape}
    \end{equation}
    The eigenvalue equation
    \begin{equation}
    \QMap \, \psi = \ue^{-\ui \theta - \gamma/2}\,\psi
    \end{equation}
    defines resonance states $\psi$
    with decay rate $\gamma$,
    i.e., their norm decays in each application
    of $\QMap$ by a factor of $\ue^{-\gamma}$.
    In autonomous scattering the phase
    $\theta$ is related to the energy
    and $\gamma$ to the width of resonance poles
    \cite{FyoSom1997,Sch2013b}.
    Note that due to the non-unitarity of $\QMap$
    the set of all right eigenfunctions
    $\{\psi_k\}_{k=1}^N$ is generally non-orthogonal,
    $\braket{\psi_k}{\psi_l} \neq 0$ for
    $k\neq l$.
    Together with the eigenfunctions $\{\phi_j\}_{j=1}^N$ of the adjoint
    $\QMap^\dagger$, called left eigenfunctions,
    a dual basis of the Hilbert space is formed.
    Without loss of generality we investigate the intensity statistics of
    right eigenfunctions in the following.
    Note that the fixed reflectivity function $\refl(\xbold)$ implies
    that in the semiclassical limit, $N\rightarrow \infty$, escape takes
    place from a large region compared to the Planck cell $h$.

    As an illustrative example we choose a paradigmatic
    two-dimensional chaotic
    map, the standard map with partial escape, as defined in
    App.~\ref{SEC:StandardMap}.
    For this system we show numerical results in Fig.~\ref{FIG:ScaledHusimis}.
    The distribution of decay rates $\gamma$,
    see Fig.~\ref{FIG:ScaledHusimis}~(a),
    extends approximately from
    the natural to the inverse decay rate of the classical map,
    $\gnat \lesssim \gamma \lesssim \ginv$,
    \cite{ClaAltBaeKet2019} and is peaked around the so-called
    typical decay rate $\gtyp$ of classically ergodic orbits~\cite{NonSch2008}.

    For the intensity statistics of
    resonance states $\psi$ we use as an example
    the Husimi function $\hus{\psi}$ \cite{Hus1940},
    which is a smooth probability distribution on phase space.
    It is defined using the overlap of the state $\psi$ with a
    coherent state $\alpha(\xbold)$ centered at some phase-space
    point $\xbold=(q,p)$,
    \begin{equation}
    \hus{\psi}(\xbold) = h^{-1} \; | \braket{\alpha(\xbold)}{\psi} |^2 .
    \end{equation}
    In the following numerical illustrations the width of
    $\alpha(\xbold)$ is chosen to be symmetric in phase space.
    Figure~\ref{FIG:ScaledHusimis}~(b) shows the Husimi function
    for three resonances states with different decay rates $\gamma$.
    They fluctuate by many orders of
    magnitude on the scale of  Planck's constant $h$.
    Their intensities clearly
    depend on the phase-space region and decay rate.
    The intensity statistics over all resonance states 
    gives a non-universal, system specific distribution.
    Even at a single phase-space point one finds
    a non-universal distribution (not shown).
    It turns out to be essential to consider
    the strong $\gamma$-dependence of resonance states. 
    Their phase-space structure
    changes significantly with $\gamma$ from an orientation
    along the classical unstable direction (close to $\gnat$)
    to the stable direction (close to $\ginv$) \cite{ClaAltBaeKet2019}.
    This is prominently seen in
    the average Husimi function $\aHg(\xbold)$ in Fig.~\ref{FIG:ScaledHusimis}~(c),
    which is defined as an  average over $\navg$ Husimi functions with decay rates close
    to the given decay rate $\gamma$.
    Note that this average structure is understood approximately by the classical dynamics
    \cite{ClaAltBaeKet2019},
    with deviations in the semiclassical limit for most $\gamma$.
    We will use numerically determined averages 
    $\aHg(\xbold)$
    when analyzing intensity fluctuations in the following.

    Comparing Figs.~\ref{FIG:ScaledHusimis}~(b) and (c)
    one observes that in regions with larger average values
    the individual Husimi functions show larger fluctuations.
    In order to obtain universality, this
    suggests to define the scaled Husimi function
    \begin{equation}
    \tHpsi(\xbold) = \frac{\hus{\psi}(\xbold)}{\aHg(\xbold)} ,
    \label{EQ:ScaledHusimi}
    \end{equation}
    which uses
    the average Husimi function $\aHg$ for scaling
    where $\gamma$ is the decay rate of $\psi$.
    The scaled Husimi functions $\tHpsi(\xbold)$
    appear universal for all $\gamma$,
    showing uniform fluctuations on phase-space,
    see Fig.~\ref{FIG:ScaledHusimis}~(d).
    This resembles the uniformity of
    eigenfunctions of closed chaotic
    quantum maps \cite{NonVor1998,Bae2003}.

    For the distribution of scaled intensities we present
    a conjecture in the next section. Subsequently, we
    numerically investigate the statistics and present numerical
    support for the conjecture in Sec.~\ref{SEC:NumericalResults}.
    
    \section{Conjecture on universal intensity statistics}
    \label{SEC:Conjecture}

    We consider the intensity statistics
    of resonance states $\psi$
    with respect to some arbitrary quantum state $\varphi$,
    i.e., the intensities
    \begin{equation}
    \Iphi := | \braket{\varphi}{\psi} |^2.
    \label{EQ:IPhi}
    \end{equation}
    Specific examples are $\varphi = \alpha(\xbold)$
    being a coherent state
    giving the Husimi function,
    $I_{\alpha(\xbold)}(\psi)  \propto \hus{\psi}(\xbold)$, and
    $\varphi = q$ being a position eigenstate giving
    the intensity in position space, $I_{q}(\psi) = |\braket{q}{\psi}|^2$.

    We conjecture for resonance states $\psi$ in chaotic systems with escape
    that the intensities
    $\Iphi$ are exponentially distributed
    with mean value $\meanI$, depending on the considered $\varphi$ and
    the decay rate $\gamma$ of $\psi$.
    Equivalently, the scaled intensities
    \begin{equation}
    \tilIphi = \frac{\Iphi}{\meanI}
    \label{EQ:ScaledIntensity}
    \end{equation}
    are exponentially distributed with mean one,
    \begin{equation}
    P[\tilIphi = {w}]\ \ud {w}  = \ue^{-{w}}\ \ud {w}.
    \label{EQ:ConjectureI}
    \end{equation}
    In particular this means that the statistics of the scaled intensities
    $\tilIphi$ is universal,
    i.e., independent of the choice of system, $\varphi$, and $\gamma$.
    Thus $\tilIphi$ shows the same statistics as intensities of eigenfunctions
    of closed chaotic
    quantum systems, where an intuitive understanding is well established in terms of a random wave model
    \cite{Ber1977b}.
    Note that similarly one could conjecture a Gaussian distribution
    of the complex amplitudes $\braket{\varphi}{\psi}$.
    We discuss a corresponding random vector model
    for systems with escape in App.~\ref{SEC:RandomVectorModel},
    which describes the statistics of $\psi$ based on an assumption in one
    distinguished basis.

    We emphasize that the mean values $\meanI$ are essential for scaling,
    but they are non-trivial compared to closed systems, where ergodicity implies a uniform mean.
    %
    It should be possible
    to relate $\meanI$ to a semiclassical limit measure for each decay rate $\gamma$,
    in particular if $\varphi = \alpha(\xbold)$ is a coherent state on
    phase space.
    However, typically these semiclassical limit measures
    are just approximately known~\cite{ClaAltBaeKet2019}
    and are multifractal without smooth phase-space densities.

    \begin{figure}[b!]
        \includegraphics[scale=1.0]{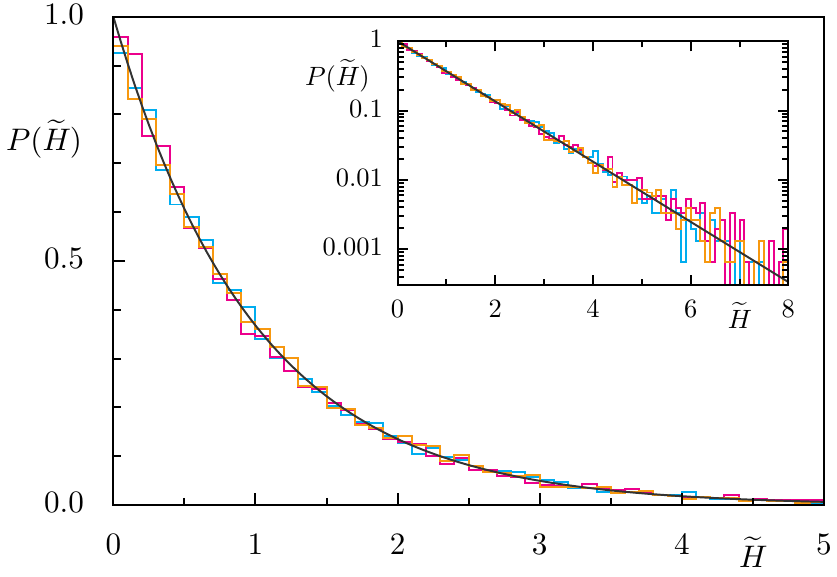}
        \caption{Distribution of scaled Husimi functions $\tHpsi$ at three
            phase-space
            points $\xbold = (q, p) \in \{(0.1, 0.2), (0.5, 0.5), (0.8, 0.7)\}$
            using $\navg = 200$ and $1/h = 16000$.
            Resonance states with decay rates $\gamma \in [\gnat, \ginv - 0.1]$ are used.
            The exponential distribution with mean one,
            Eq.~\eqref{EQ:ConjectureI},
            is shown as a black line.
            The inset shows the comparison on a semi-logarithmic scale.
            The system is the chaotic standard map with partial escape,
            see App.~\ref{SEC:StandardMap}.
        }
        \label{FIG:StatisticsSinglePoint}
    \end{figure}
    \section{Numerical results}
    \label{SEC:NumericalResults}
    
    In the following we present numerical support for the above conjecture
    on intensity statistics for systems with escape.
    This is done in Sec.~\ref{SEC:NumericalResultsStandard}
    for the standard map
    and in Sec.~\ref{SEC:NumericalResultsBaker} for the triadic baker map
    with partial escape,
    where in both cases
    the mean $\meanI$ is numerically approximated.
    In Sec.~\ref{SEC:NumericalResultsRandom}
    a random matrix model is considered,
    where the mean $\meanI$ is analytically determined.

    \subsection{Standard map with partial escape}
    \label{SEC:NumericalResultsStandard}
    
    For the numerical approximation
    of the mean intensity $\meanI$ it is
    necessary to define which resonance states are
    used for averaging.
    On one hand they should have approximately the same decay rate
    due to the strong dependence of their structure on $\gamma$,
    see Fig.~\ref{FIG:ScaledHusimis}~(b).
    On the other hand the number $\navg$ of considered states
    should be large enough, such that the average is not too much
    distorted by fluctuations of individual states.
    This finite sample effect would lead to deviations from the exponential distribution,
    see App.~\ref{SEC:FiniteSize}.
    In order to scale a state $\psi$ with decay rate $\gamma$,
    we choose $\navg = 200$  for $1/h=16000$,
    selecting the closest $\navg/2$ states with decay rates
    greater and smaller than $\gamma$, respectively.
    \begin{figure}[t!]
        \includegraphics[scale=1.]{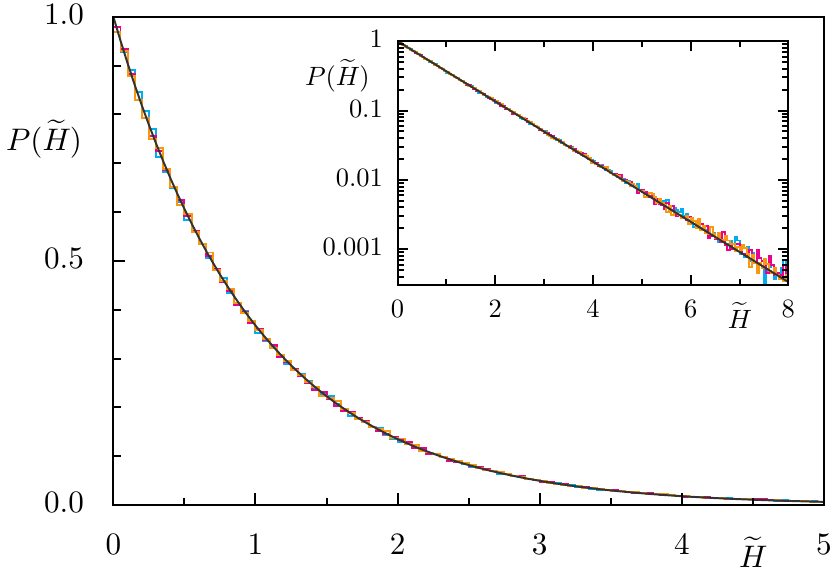}
        \caption{Distribution of scaled Husimi functions $\tHpsi$
            close to three decay rates
            $\gnat \approx 0.22$, $\gtyp \approx 0.48$, and $\ginv - 0.1 \approx 0.78$
            for $\nsample = 300$ resonance states each,
            evaluated on a $50\times50$ phase-space grid
            using $\navg = 200$ and $1/h = 16000$.
            The exponential distribution with mean one,
            Eq.~\eqref{EQ:ConjectureI},
            is shown as a black line.
            The inset shows the comparison on a semi-logarithmic scale.
            The system is the chaotic standard map with partial escape,
            see App.~\ref{SEC:StandardMap}.
        }
        \label{FIG:StatisticsAll}
    \end{figure}
    \begin{figure}[b!]
        \includegraphics[scale=1.]{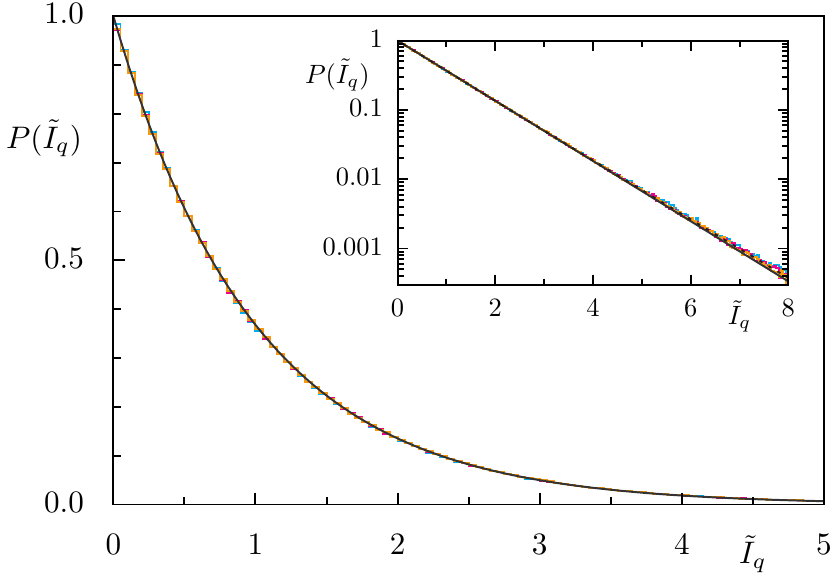}
        \caption{Distribution of scaled position intensities $\tilIphi[q]$ for all
            $16000$ positions $q$,
            using the same parameters as in Fig.~\ref{FIG:StatisticsAll}.
            The dashed line in the inset shows additionally
            the expected distribution when scaling with an average
            for finite $\navg = 200$,
            Eq.~\eqref{eq:corrected_pdf}.
        }
        \label{FIG:StatisticsPosBasis}
    \end{figure}
    \begin{figure}[t!]
        \includegraphics[scale=1.0]{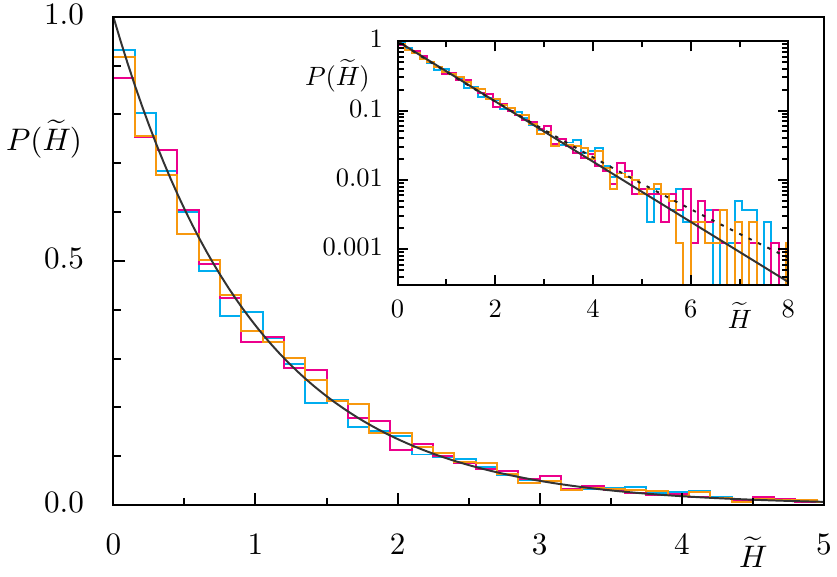}
        \caption{Distribution of scaled Husimi functions $\tHpsi$
            for $h=1/250$, $\navg = \nsample = 24$, and considering
            a $15\times 15$ phase-space grid,
            otherwise as in Fig.~\ref{FIG:StatisticsAll}.
            The dashed line in the inset shows additionally
            the expected distribution when scaling with an average
            for finite $\navg = 24$,
            Eq.~\eqref{eq:corrected_pdf}.
        }
        \label{FIG:StatisticsHusimiQuantumRegime}
    \end{figure}
    
    For the standard map with escape we focus on the
    intensity statistics of the Husimi function.
    %
    We first consider the distribution for fixed phase-space
    points $\xbold$ using a large range of decay rates $\gamma$.
    Secondly, we investigate the statistics fixing $\gamma$ and
    considering many phase-space points.
    Note that
    fixing both $\gamma$ and $\xbold$ would lead to much less
    intensity values and thus be insufficient to critically test for the
    exponential distribution.

    First, we investigate the scaled Husimi functions $\tHpsi$
    for three phase-space points $\xbold$.
    For this purpose we consider all resonance states $\psi$
    from the natural decay rate up to close to the inverse decay rate.
    Note that we exclude $\navg/2$ resonance states at each end
    of the decay rate distribution,
    for which the number of surrounding resonance states
    is not sufficient for the computation of the average.
    The intensity distribution is shown in
    Fig.~\ref{FIG:StatisticsSinglePoint}.
    It nicely follows the exponential distribution,
    Eq.~\eqref{EQ:ConjectureI},
    with no statistically significant
    deviations for all three chosen phase-space points,
    supporting the conjecture.

    Secondly, we consider for each of the
    three decay rates $\gnat$, $\gtyp$, and $\ginv-0.1$
    a sample of $\nsample = 300$ scaled Husimi functions
    with close-by decay rates $\gamma$.
    They are calculated on a phase-space grid of size $50\times 50$,
    which for $1/h = 16000$ has negligible intensity correlations of
    neighboring points.
    Their intensity distribution is shown in Fig.~\ref{FIG:StatisticsAll}.
    It follows the exponential distribution,
    Eq.~\eqref{EQ:ConjectureI},
    with no statistically significant
    deviations for all three decay rates,
    again supporting the conjecture.

    In order to validate the conjecture in a different basis
    we illustrate the distribution of scaled position intensities
    $\tilIphi[q]$
    in Fig.~\ref{FIG:StatisticsPosBasis},
    where the same parameters as in Fig.~\ref{FIG:StatisticsAll} are used.
    The intensity distribution nicely follows the exponential distribution,
    Eq.~\eqref{EQ:ConjectureI}.
    The inset reveals for large $\tilIphi[q] > 6$ a small systematic
    deviation,
    which we attribute to scaling by an average using finite
    $\navg$, see App.~\ref{SEC:FiniteSize}.
    This systematic deviation is visible, 
    as the fluctuations are smaller than in Fig.~\ref{FIG:StatisticsAll}.

    Let us finally investigate if the
    intensity statistics follows the conjecture in the
    quantum regime for larger $h$.
    For this we consider scaled Husimi functions $\tHpsi$ of
    resonance states for $h=1/250$.
    Since the number of resonance states is much smaller in
    sufficiently small $\gamma$-intervals, 
    we use $\navg = 24$ and $\nsample = 24$.
    The intensity distribution is shown in Fig.~\ref{FIG:StatisticsHusimiQuantumRegime}.
    It nicely follows the exponential distribution,
    Eq.~\eqref{EQ:ConjectureI}, with larger fluctuations
    than for $h=1/16000$, as expected. 
    For $\tilIphi[q] > 6$ the distribution deviates from the exponential
    behavior, which we attribute to scaling by an average using finite
    $\navg$, see App.~\ref{SEC:FiniteSize}.
    Thus even towards the quantum regime we find support of the conjecture.

    Let us mention that we find numerical support for
    the conjecture for several reflectivity functions,
    by varying the region and strength of escape,
    see App.~\ref{SEC:StandardMap}.
    We also confirmed the conjecture
    for the case of a region with full escape,
    where the semiclassical support of
    resonance states converges to the so-called backward trapped set
    \cite{KeaNovPraSie2006}, such that
    one has to restrict the analysis to phase-space points in this set (not shown).

    \subsection{Baker map with partial escape}
    \label{SEC:NumericalResultsBaker}
    \begin{figure}[t!]
        \centering
        \includegraphics[scale=1.]{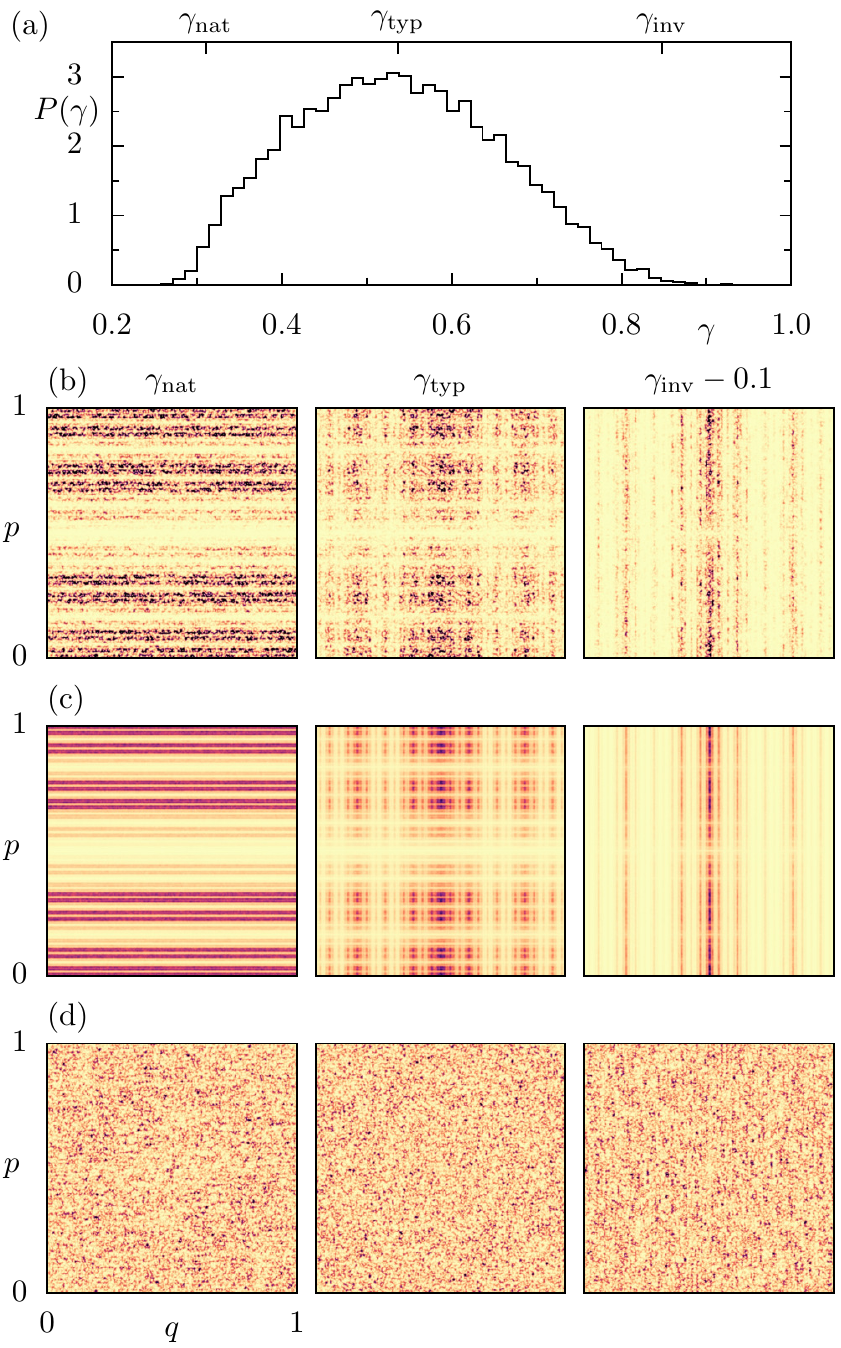}
        \caption{Same as Fig.~\ref{FIG:ScaledHusimis} for the triadic baker map
            with partial escape,
            defined in App.~\ref{SEC:BakerMap}, using $h=1/16002$.
            The classical decay rates are
            $\gnat \approx 0.31$, $\gtyp \approx 0.54$, and $\ginv \approx 0.85$.
        }
        \label{FIG:ScaledHusimisBaker}
    \end{figure}
    \begin{figure}[t!]
        \centering
        \includegraphics[scale=1.]{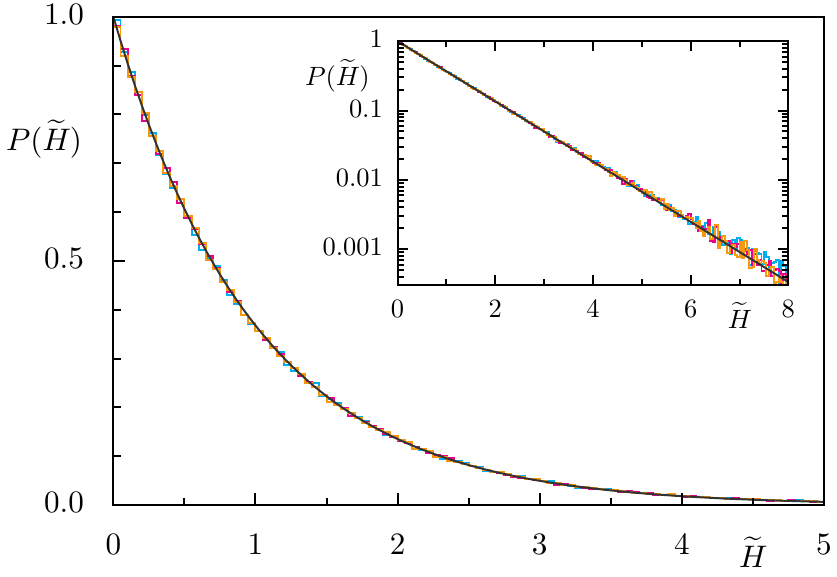}
        
        \caption{Distribution of scaled Husimi functions $\tHpsi$,
            for the triadic baker map with partial escape,
            defined in App.~\ref{SEC:BakerMap}, using $h=1/16002$
            and considering a $60\times60$ phase-space grid.
            Otherwise parameters as in Fig.~\ref{FIG:StatisticsAll}.
            The considered decay rates are $\gnat \approx 0.31$,
            $\gtyp \approx 0.54$, and
            $\ginv -0.1 \approx 0.75$.
        }
        \label{FIG:StatisticsAllBaker}
    \end{figure}

    In this section we present results for the baker map with escape
    which is a well-studied model for chaotic resonances
    \cite{KeaNovPraSie2006,NonRub2007,KeaNonNovSie2008,%
        NovPedWisCarKea2009,PedWisCarNov2012,CarBenBor2016},
    defined in App.~\ref{SEC:BakerMap}.
    The classical baker map is
    ergodic, uniformly hyperbolic, %
    and explicit expressions for all periodic orbits are available.
    For the mean intensity of resonance states
    a sufficiently accurate classical description is not known
    and we use numerical approximations, as for the standard map.

    In Fig.~\ref{FIG:ScaledHusimisBaker} we show single, average and scaled
    Husimi functions for three different decay rates $\gamma$ for the triadic baker map
    with partial escape at $h = 1/16002$.
    The single and average Husimi functions reveal a structural change with increasing
    $\gamma$
    from extending along
    the classical unstable $q$-direction to the
    stable $p$-direction \cite{ClaAltBaeKet2019}.
    The scaled Husimi functions fluctuate uniformly
    on phase space.
    Their intensity statistics is shown in Fig.~\ref{FIG:StatisticsAllBaker}
    and follows the exponential distribution,
    Eq.~\eqref{EQ:ConjectureI},
    with no statistically significant
    deviations for all three decay rates,
    again supporting the conjecture.
    Similar results for an asymmetric baker map with partial escape
    are shown in App.~\ref{SEC:BakerMap}.

    \subsection{Random matrix model with partial escape}
    \label{SEC:NumericalResultsRandom}
    
    In this section we introduce a random matrix model with escape
    and numerically support the conjecture.
    The motivation for this model is
    (i)
    that the mean intensity $\meanI$ can be described analytically
    and
    (ii)
    that it should allow for a rigorous proof of the conjecture.
    Note, that the phase-space distribution of resonance states in this
    model is not multifractal.

    We replace the propagator $\QMapcls$ in Eq.~\eqref{EQ:QMapEscape} with
    a random matrix $\MMapcls$ of dimension $N$ taken from the circular unitary ensemble
    \cite{FyoSom2000,ZycSom2000}.
    The eigenstates of
    $\MMap = \MMapcls\Qrefl$ have a much simpler phase-space structure than in
    systems with deterministic dynamics,
    depending on the decay rate $\gamma$
    and the reflectivity function $r(\xbold)$.
    For such a system the mean intensities
    $\meanI[\alpha(\xbold)]$ on phase space or
    $\meanI[q]$ in position space are given by classical
    densities $\rho_\gamma(\xbold)$ and $\rho_\gamma(q)$, respectively.
    A derivation of these densities is given
    in App.~\ref{SEC:ClassicalDensitiesRMMwE}.

    \begin{figure}[t!]
        \centering
        \includegraphics[scale=1.]{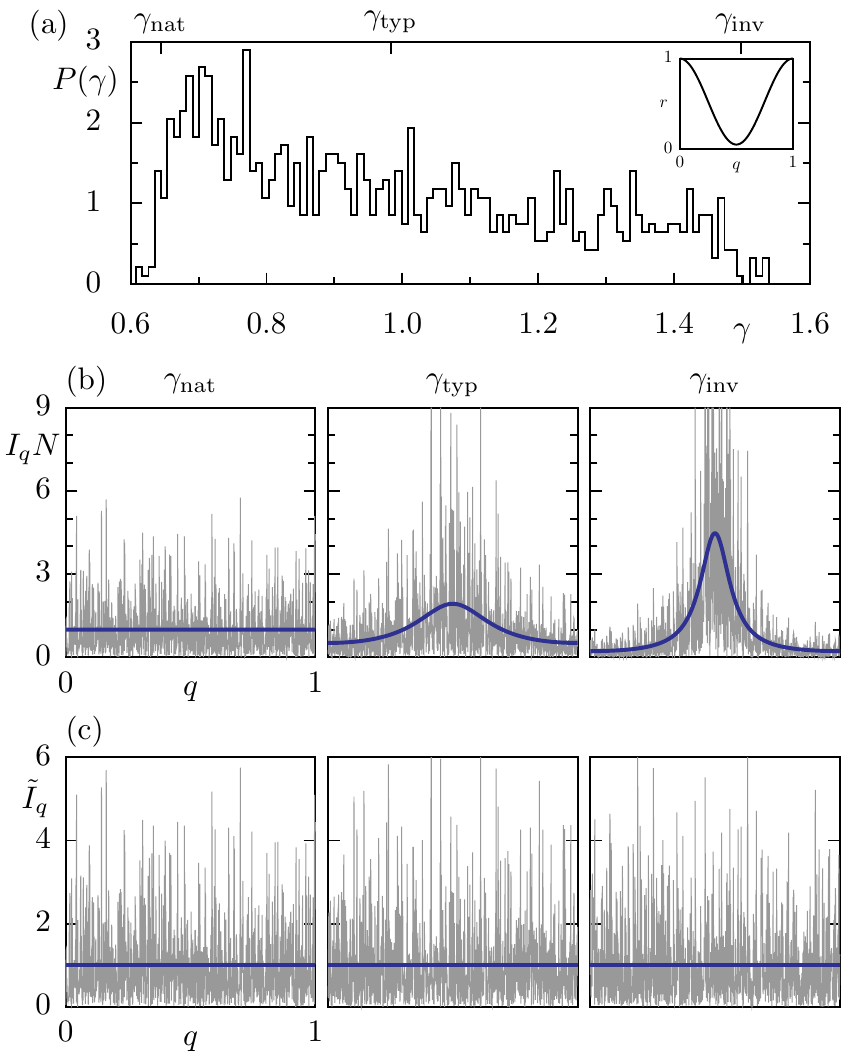}
        \caption{
            (a) Distribution of quantum decay rates $\gamma$
            for a random matrix with partial escape using
            $N = 1000$.
            The inset shows the considered reflectivity function $r(q)$.
            The classical decay rates
            $\gnat\approx0.64, \gtyp\approx0.98, \ginv\approx1.5$ are
            indicated.
            (b)
            Intensities in position space $I_q(\psi)N$
            for three exemplary resonance states $\psi$ with decay rates
            $\gnat$, $\gtyp$, and $\ginv$ and
            classical density $\rhoxi(q)$,
            Eq.~\eqref{EQ:RMM_Limit} (thick line).
            (c) Scaled intensities $\tilde{I}_q(\psi)$
            for the same decay rates, compared to the uniform density
            (thick line).%
        }
        \label{FIG:RandomModelPosition}
    \end{figure}
    \begin{figure}[t!]
        \includegraphics[scale=1.0]{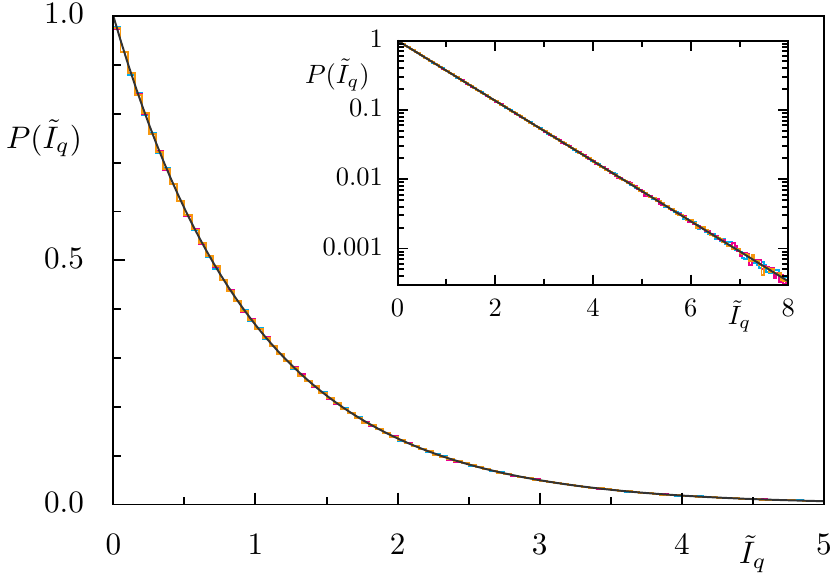}
        \caption{%
            Distribution of scaled intensities $\tilIphi[q]$
            close to the three decay rates
            $\gnat \approx 0.64$,
            $\gtyp \approx 0.98$, and
            $\ginv \approx 1.5$
            for $\nsample = 300$ resonance states each,
            evaluated for all position states $q$
            for $N = 16000$ using $\rhoxi(q)$, Eq.~\eqref{EQ:RMM_Limit},
            for scaling.
            The exponential distribution with mean one,
            Eq.~\eqref{EQ:ConjectureI},
            is shown as a black line.
            The inset shows the comparison on a semi-logarithmic scale.
            The system is the random matrix model with partial escape.
        }
        \label{FIG:RandomMatrixFluctuationsLargeMatrix}
    \end{figure}%
    Numerical results are
    presented for an exemplary smooth reflectivity function,
    $\refl(q, p) = 1 - (1 - \alpha)\sin^2(\pi q)$ with $\alpha = 0.05$,
    see inset of Fig.~\ref{FIG:RandomModelPosition}~(a).
    The distribution of decay rates $\gamma$, see Fig.~\ref{FIG:RandomModelPosition}~(a),
    extends approximately from
    the natural to the inverse decay rate.
    The intensity in position representation
    of resonance states for three different decay rates
    is shown in
    Fig.~\ref{FIG:RandomModelPosition}~(b).
    These intensities fluctuate around the smooth classical densities
    $\rhoxi$, shown for comparison.
    Averaging over several resonance states with close-by decay rates
    (or local averaging in position space)
    we find for large matrix dimension $N$
    perfect agreement with the classical density
    for the natural and inverse natural decay
    as well as for decay rates in between (not shown).
    We scale the intensities $\Iphi[q]$
    with the classical densities, i.e.,
    using $\meanI[q] = \rhoxi(q) / N$.
    The scaled intensity $\tilIphi[q]$ seems to be
    independent of the position $q$ and decay rate $\gamma$,
    see Fig.~\ref{FIG:RandomModelPosition}~(c),
    illustrating the universality.
    The conjectured
    exponential distribution, Eq.~\eqref{EQ:ConjectureI}, is validated
    in Fig.~\ref{FIG:RandomMatrixFluctuationsLargeMatrix}.
    
    %
    
    \section{Conclusion and outlook}
    \label{SEC:Conclusion}
    
    In summary, we conjecture that the fluctuations of scaled intensities for
    resonance states in chaotic quantum
    systems with escape are universally described
    by the exponential distribution with mean one.
    This generalizes well-known results for closed chaotic systems.
    Numerically we investigate the statistics of single resonance states
    for the chaotic standard map and the triadic baker map,
    which are  suitably scaled by their respective $\gamma$-dependent
    multifractal average.
    The Husimi statistics for all considered cases of different phase-space points and decay rates
    agrees excellently
    with the conjectured exponential distribution.
    We demonstrate the conjecture in position basis,
    deep in the quantum regime for large $h$, and for different reflectivity
    functions $\refl$.
    This is further confirmed in
    a random matrix model with partial escape,
    for which the semiclassical limit densities of resonance states are
    derived analytically.

    Generic dynamical systems are not fully chaotic, but instead 
    regions of regular and chaotic dynamics coexist.
    The important question arises whether the conjecture applies to such mixed 
    systems with escape.
    Indeed we find in a preliminary study 
    that the intensity statistics follows the conjecture, 
    if just the subset of chaotic resonance states is considered
    and if it is investigated just on the chaotic region, 
    while we find no universal statistics for regular resonance states.

    The presented analysis does not show signatures of
    scarring of individual resonances on periodic orbits
    \cite{WisCar2008}.
    In fact, all locally enhanced intensities are consistent with
    enhancements of the average (which is a multifractal
    of classical origin) and exponentially distributed fluctuations
    on top of that.
    Since the presented statistics involves many resonances,
    signatures of scarring of individual resonances
    are possibly concealed.
    It would be interesting to study the relation
    between scarring and the observed universal statistics in the future.

    There are various further directions in which these results can be generalized:
    (i)
    It is interesting to understand if and how different symmetry
    classes of the closed map and symmetries of the reflectivity function
    lead to different intensity statistics.
    (ii) We speculate that the
    autocorrelation function of scaled Husimi functions
    behaves as in closed systems \cite{NonVor1998,Sch2005b}.
    (iii) Another interesting aspect are implications of these results on
    the fluctuations in the near- and far-field emission of optical microcavities,
    which are experimentally accessible \cite{BitKimZenWanCao2020}.
    %

    \acknowledgments
    
    We thank J.~Keating, S.~Nonnenmacher, M.~Novaes, S.~Prado, and M.~Sieber
    for helpful discussions.
    This research is
    funded by the Deutsche Forschungsgemeinschaft
    (DFG, German Research Foundation) - 262765445.
    
    \begin{figure}[b!]
        \includegraphics[scale=1.]{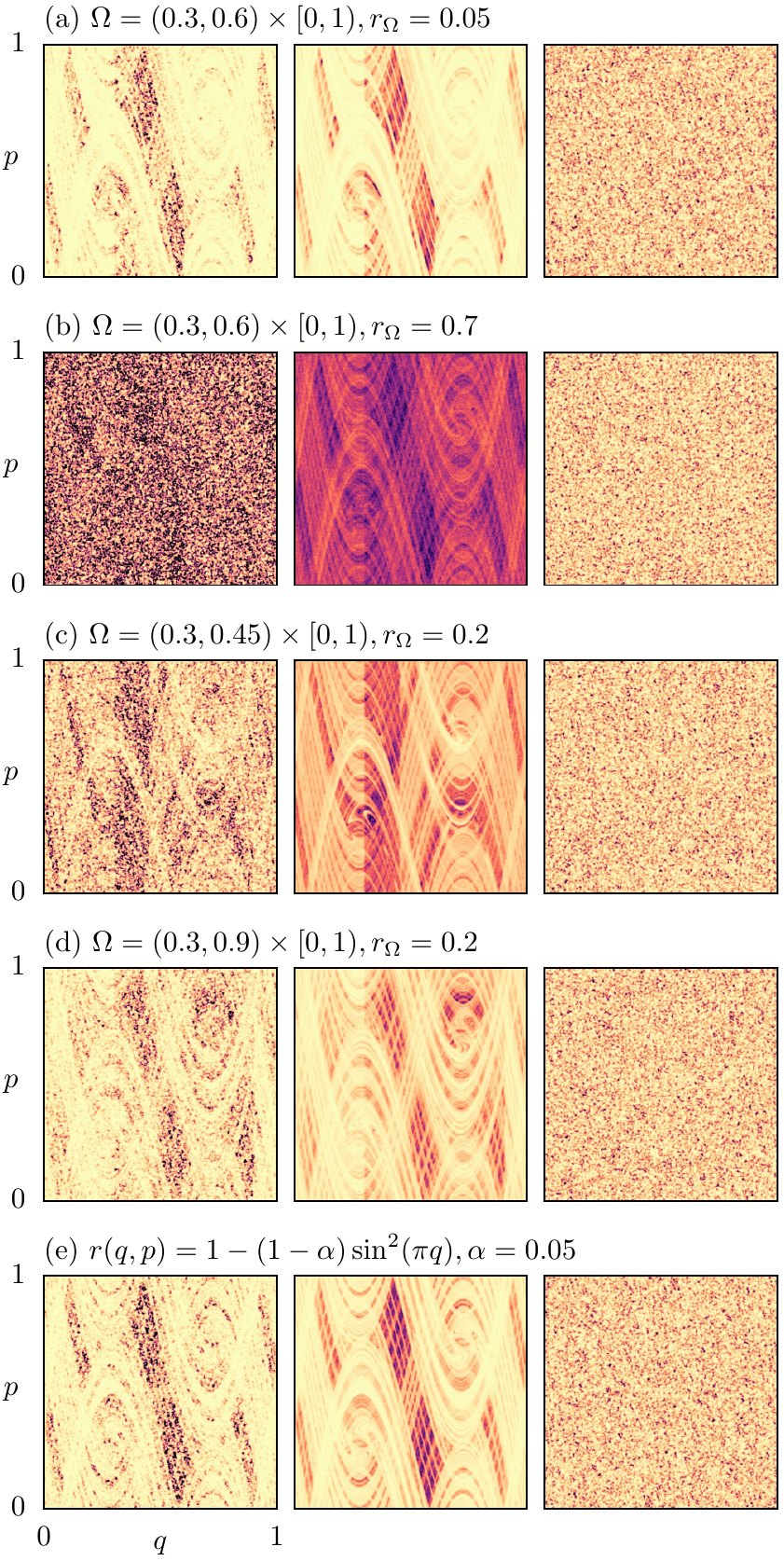}
        \caption{Single $\hus{\psi}$, averaged $\aHg$, and scaled $\tHpsi$ 
            Husimi functions (left to right)
            for resonance states $\psi$ with decay rate closest to
            $\gtyp$ for the chaotic standard map with different $\refl(q, p)$,
            using $h=1/16000$ and $\navg = 200$.
            The typical decay rates are 
            (a) $\gtyp \approx 0.90$,
            (b) $\gtyp \approx 0.11$,
            (c) $\gtyp \approx 0.24$,
            (d) $\gtyp \approx 0.97$, and
            (e) $\gtyp \approx 0.98$.
            Colormaps as in Fig.~\ref{FIG:ScaledHusimis}.
        }
        \label{FIG:ScaledHusimisAppendix}
    \end{figure}
    \appendix
    
    \section{Standard map with escape}
    \label{SEC:StandardMap}

    \begin{figure}[b!]
        \includegraphics[scale=1.]{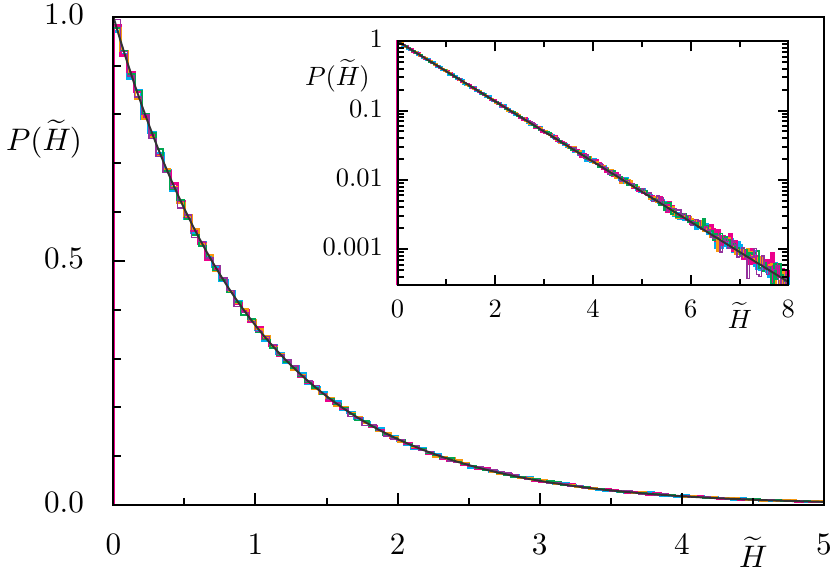}
        \caption{Distribution of scaled Husimi functions $\tHpsi$
            with decay rates close to $\gtyp$
            for the chaotic standard map with different $\refl(q,p)$,
            as specified in Fig.~\ref{FIG:ScaledHusimisAppendix}.
            Other parameters as in Fig.~\ref{FIG:StatisticsAll}.
        }
        \label{FIG:StatisticsAllAppendix}
    \end{figure}
    As an example system we consider the paradigmatic standard map
    on the torus
    \cite{Chi1979}, which is given by
    the time-periodically driven Hamiltonian
    $H(q, p, t) = p^2/2 + \sum_{n=-\infty}^\infty V(q)\delta(t - n)$
    with kicking potential $V(q) = \kappa/(4\pi^2)\cos(2\pi q)$ and
    dimensionless coordinates $(q, p)\in [0,1)\times[0,1)$.
    We consider the half-kick map
    $\Mapcls(q, p) = (q + p^\ast, p - V'(q + p^\ast)/2)$
    with
    $p^\ast = p - V'(q)/2$
    (similar results are expected for other variants of the map).
    For $\kappa = 10$ the phase-space contains
    no visible regular regions, such that we call this setting
    the chaotic standard map.
    One possible quantization of this map
    is given by the unitary propagator between two kicks,
    which is determined by Floquet quantization
    \cite{BerBalTabVor1979,ChaShi1986}. For the half-
    kick mapping it reads
    \begin{equation}
    \QMapcls = \ue^{-\ui/(2\hbar)V(q)}\ue^{-\ui/(2\hbar)p^2}
    \ue^{-\ui/(2\hbar) V(q)},
    \end{equation}
    where $h = 2\pi\hbar$ takes the role of an effective Planck constant
    due to dimensionless units $q$ and $p$. Considering periodic
    boundary conditions, i.e., dynamics on a torus, only discrete
    values $h = 1/N$ with $N \in\mathbb{N}$ are allowed. The semiclassical
    limit is described by $h \rightarrow 0$.
    
    We consider partial escape through some region $\Omega$, such
    that the reflectivity function is given by
    $\refl(q,p)=\reflOmega < 1$ for $(q,p)\in\Omega$ and
    $\refl(q,p) = 1$ for $(q,p)\notin \Omega$.
    This leads to a projective coupling operator \cite{NonSch2008} of the form
    $\Qrefl = P_{\Omega^c} + \sqrt{\reflOmega} P_\Omega$.
    In the main text we use $\Omega = (0.3, 0.6)\times [0,1)$ and
    $\reflOmega = 0.2$.

    Here we present additional results for different
    reflectivity functions $\refl(q, p)$.
    For this purpose we first consider the same opening
    $\Omega$ and choose two different reflectivities $\reflOmega$ leading to
    much stronger and weaker escape from the system, respectively.
    Secondly, we choose smaller and larger openings $\Omega$ for the same
    reflectivity $\reflOmega$.
    Finally, also a smooth reflectivity function is considered.
    In Fig.~\ref{FIG:ScaledHusimisAppendix}
    for these five different choices of $\refl(q, p)$ the 
    single, averaged, and scaled Husimi functions for the respective
    decay rate $\gtyp$ are shown.
    In all cases the scaled Husimi function
    is uniform on phase space (rightmost panels).
    The corresponding intensity statistics 
    is shown in Fig.~\ref{FIG:StatisticsAllAppendix}.
    For all considered reflectivity functions
    it nicely follows the conjectured exponential distribution
    with no statistically significant deviations.

    \begin{figure}[b!]
        \centering
        \includegraphics[scale=1.]{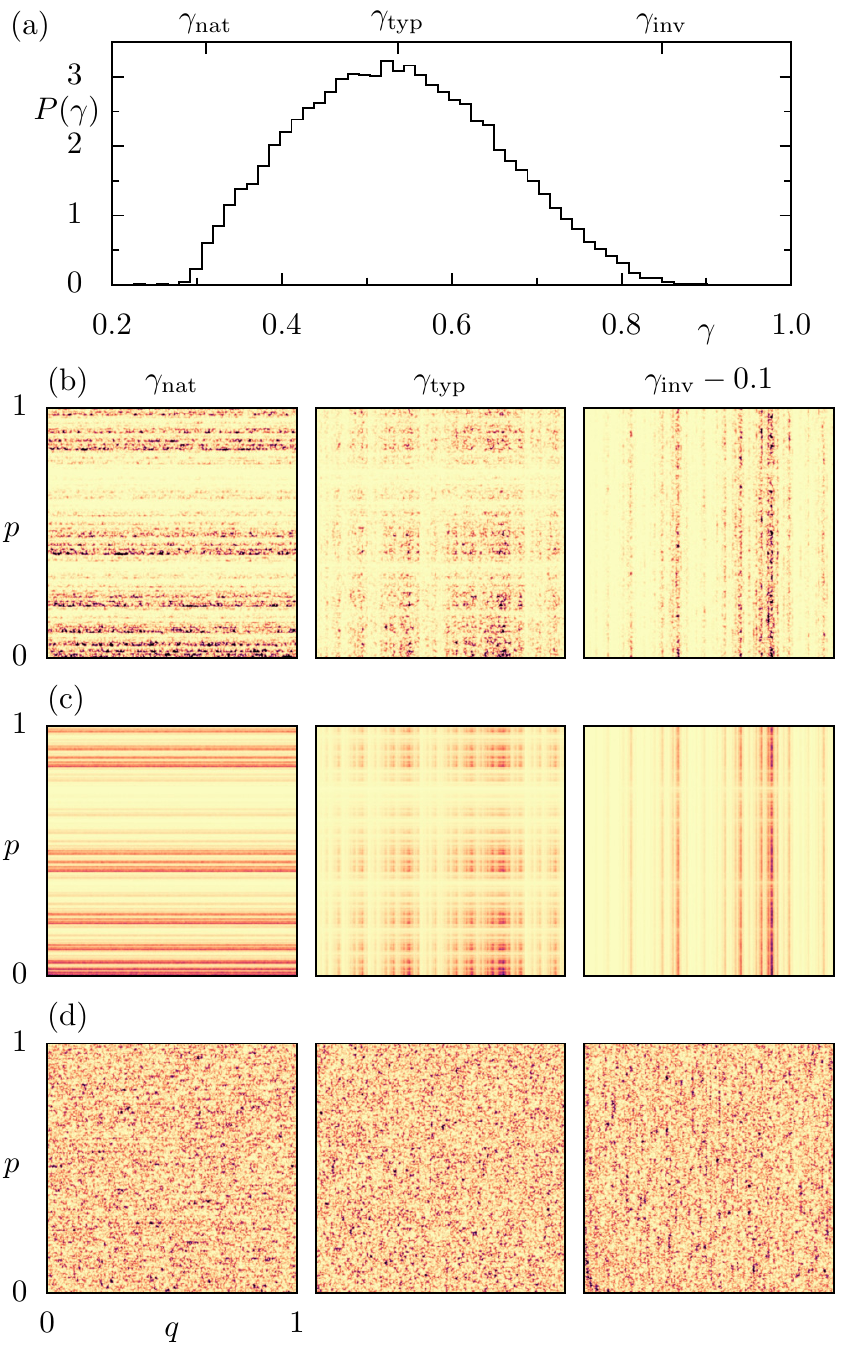}
        \caption{Same as Fig.~\ref{FIG:ScaledHusimis} for the asymmetric baker map
            with partial escape using $h=1/16002$.
            The classical decay rates are
            $\gnat \approx 0.31$, $\gtyp \approx 0.54$, and $\ginv \approx 0.85$.
        }
        \label{FIG:ScaledHusimisBakerII}
    \end{figure}
    \section{Baker map with escape}
    \label{SEC:BakerMap}
    
    \begin{figure}[t!]
        \centering%
        \includegraphics[scale=1.]{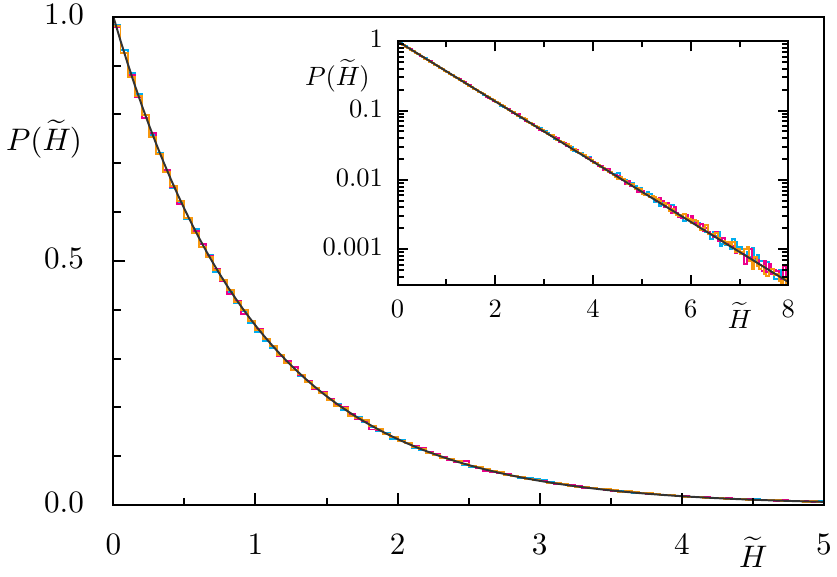}
        
        \caption{Distribution of scaled Husimi functions $\tHpsi$
            for the asymmetric baker map with partial escape
            using $h=1/16002$ and considering a $60\times60$ phase-space grid.
            Other parameters as in Fig.~\ref{FIG:StatisticsAll}.
            The considered decay rates are $\gnat \approx 0.31$,
            $\gtyp \approx 0.54$, and
            $\ginv -0.1 \approx 0.75$.
        }
        \label{FIG:StatisticsAllBakerII}
    \end{figure}
    
    The generalized $n$-baker map on the two-torus $[0,1) \times[0, 1)$ is defined as follows
    \cite{ArnAve1968}.
    Let $\bbold\in \mathbb{R}_{>0}^n$ with $\sum_{i=1}^n b_i =1$ denote
    the relative size of $n$ vertical rectangles $A_k = [a_k, a_k + b_k) \times [0,1)$,
    where $a_k := \sum_{i=1}^{k - 1} b_i$ corresponds to the left boundary of $A_k$
    (with $a_1 = 0$ and $a_{k + 1} = 1$).
    With this, the baker map is defined as
    $\BakerMap ( q, p) := ((q - a_k) / b_k, b_k p + a_k)$ for $q \in [a_k, a_{k+1})$.
    I.e., in one step the $i$-th rectangle is compressed along the $p$-direction by the factor $b_i$ and stretched
    along the $q$-direction by the factor $1/b_i$, after which
    being stacked on top of each other.
    The quantized baker map is given by \cite{BalVor1989,Sar1990}
    \begin{equation}
    \QBakerMap = \mathcal{F}_N^{-1}
    \operatorname{diag}(\mathcal{F}_{N_1}, \mathcal{F}_{N_2}, \dots, \mathcal{F}_{N_n}),
    \end{equation}
    where $[\mathcal{F_M}]_{kl} :=
    M^{-1/2}\ue^{-2\pi\ui (k+1/2)(l+1/2) / M}$
    denotes the discrete Fourier transform of dimension $M$,
    and $N_i / N = b_i$.

    For the baker map with escape
    we consider reflectivity functions which are constant in each rectangle,
    $\refl(q,p) = \refl_k$ for $q \in [a_k, a_{k+1})$ for some
    $\reflbold \in \mathbb{R}_{\geq0}^n$.
    This allows to determine the classical decay rates analytically, which gives
    $\gnat = -\ln \sum_{i=1}^n b_i \refl_i$,
    $\ginv =  \ln \sum_{i=1}^n b_i / \refl_i$, and 
    $\gtyp = -  \sum_{i=1}^n b_i \ln \refl_i$.
    In Sec.~\ref{SEC:NumericalResultsBaker} results are presented for the
    triadic baker map with equal sizes $\bbold = (1/3, 1/3, 1/3)$,
    where escape from the middle strip is considered as $\reflbold = (1, 0.2, 1)$.

    Here we show additional results for an asymmetric baker map with escape, defined by
    $\bbold = (1/2, 1/3, 1/6)$ and $\reflbold = (1, 0.2, 1)$,
    see Fig.~\ref{FIG:ScaledHusimisBakerII} and Fig.~\ref{FIG:StatisticsAllBakerII}.

    \section{Deviations from exponential distribution due to
        scaling with average over finite sample}
    \label{SEC:FiniteSize}
    
    In this section we derive how the distribution
    of scaled intensities
    deviates from the exponential distribution,
    if a finite sample $\navg$ is used to determine the average.
    This becomes relevant in situations where the calculation is numerically costly,
    e.g., in billiards with escape like optical microcavities.

    Therefore, we consider intensities which are described by
    exponentially distributed random variables $X_i$ with the same mean
    ${1}/{\lambda}$. Their probability density function is then given by
    \begin{equation}
    \label{eq:exponential_pdf}
    \prob{X_i = x} = \Theta\left(x\right) \lambda \ue^{-\lambda x}
    \end{equation}
    Let $X_0$ denote the random variable that we wish to
    scale and let $n = \navg$ be the number of states contributing to the
    average. Then the random variable
    \begin{equation}
    Y = \frac{X_0}{\frac{1}{n}\sum_{i=1}^n X_i}
    \label{eq:scaled_variable}
    \end{equation}
    models the scaled intensity, see Eq.~\eqref{EQ:ScaledIntensity}.
    Its probability density function can be calculated as
    \begin{eqnarray}
    \label{eq:corrected_pdf_definition}
    \prob[n]{Y = y}
    & = &
    \int_0^{\infty} \ud x_0 \int_0^{\infty} \ud x_1 \dots
    \int_0^{\infty} \ud x_n \\ \nonumber
    && \quad \delta\left(y - \frac{ x_0}{\frac{1}{n}\sum_{i=1}^n x_i}\right)%
    \prod_{i=0}^n \prob{X_i = x_i}
    \\
    \label{eq:corrected_pdf}
    &=&
    \left(1 + \frac{y}{n}\right)^{-(n + 1)}
    \text{.}
    \end{eqnarray}
    As expected, in the limit of large $n$
    this distribution converges to the exponential distribution,
    $P_n(Y=y) \quad
    \xrightarrow{n\rightarrow\infty} \quad \ue^{-y}$.
    We observe for values of $n < 100$ that the distribution of
    scaled intensities of resonance states
    closely follows Eq.~\eqref{eq:corrected_pdf} (not shown).
    For the considered value of $\navg = 200$
    the deviation between
    Eq.~\eqref{eq:corrected_pdf} and the exponential distribution is almost not
    visible in Figs.~\ref{FIG:StatisticsSinglePoint}
    and~\ref{FIG:StatisticsAll},
    the relative error being about $4\%$ ($12\%$) for
    $y = 5$ ($y = 8$).
    Note that it is also possible to include $X_0$ in the average in
    Eq.~\eqref{eq:scaled_variable},
    leading to a slightly different distribution $P_n$,
    which converges as well to the exponential distribution for $n\rightarrow\infty$.
    
    \section{Classical densities for random matrix model with partial escape}
    \label{SEC:ClassicalDensitiesRMMwE}
    For the random matrix model with partial escape used
    in Sec.~\ref{SEC:NumericalResultsRandom}
    classical densities $\rho_{\gamma}(\xbold)$,
    which we assume to describe the mean intensities of
    resonance states, are derived as follows.
    Classically, the natural decay rate $\gnat$ and the
    natural decay rate from the inverse dynamics, $\ginv$,
    as well as their corresponding densities
    are given by
    \begin{eqnarray}
    \rhonat(\xbold)
    =&
    1,
    \qquad
    &\ue^{-\gnat}
    =
    \int \refl(\xbold)\,\ud \xbold,
    \label{EQ:RMM_nat}
    \\
    \rhoinv(\xbold)
    =&
    \displaystyle
    \frac{1}{\ue^{\ginv}\, \refl(\xbold)},
    \qquad
    &\ue^{\ginv\phantom{-}}
    =
    \int  \frac{1}{\refl(\xbold)} \, \ud \xbold.
    \label{EQ:RMM_inv}
    \end{eqnarray}
    These densities are stable under forward (backward) iteration of a
    corresponding classical time evolution. Here the (inverse)
    random matrix is replaced by a stochastic map
    which leads to the uniform density in a single step, while keeping the norm.
    Specifically, under forward iteration of $\rhonat(\xbold)$
    first the reflectivity function
    reduces the norm by the factor $\ue^{-\gnat}$
    and the random step makes the density uniform again.
    Under backward iteration of $\rhoinv(\xbold)$
    first the random step leads to the uniform density
    and then the inverted reflectivity function $1/\refl(\xbold)$
    increases the norm by the factor $\ue^{\ginv}$
    (corresponding in forward direction to a decay by $\ue^{-\ginv}$)
    and induces the phase-space density
    $\rhoinv(\xbold) \propto 1/\refl(\xbold)$.

    For arbitrary decay rates $\gamma$
    the classical density $\rhoxi(\xbold)$
    has to fulfill the condition of normalization,
    \begin{equation}
    \int \rhoxi(\xbold) \, \ud \xbold
    = 1,
    \label{EQ:RMM_norm2}
    \end{equation}
    and the condition of decay with $\gamma$ under forward iteration,
    \begin{equation}
    \int \refl(\xbold) \, \rhoxi(\xbold)\, \ud \xbold
    =
    \ue^{-\gamma}
    \int \rhoxi(\xbold) \, \ud \xbold
    .
    \end{equation}
    This can be equivalently written as
    \begin{equation}
    \int g_\gamma(\xbold)\, \rhoxi(\xbold)\, \ud \xbold = 0,
    \label{EQ:RMM_decay2}
    \end{equation}
    where 
    \begin{equation}
    g_\gamma(\xbold)
    =
    \ue^\gamma\refl(\xbold) - 1
    \label{EQ:GFunc}
    \end{equation}
    is fixed by the considered reflectivity function
    $r(\xbold)$ and the decay rate $\gamma$.
    There are in general
    infinitely many classical densities $\rho_{\gamma}(\xbold)$
    satisfying Eq.~\eqref{EQ:RMM_decay2} and
    it is not obvious which one is relevant quantum mechanically.
    We will use that due to linearity
    the solutions  $\rho_{\gamma}(\xbold)$
    of Eq.~\eqref{EQ:RMM_decay2}
    are the same 
    when replacing $g_\gamma(\xbold)$ by
    some function 
    $g(\xbold) = \xi \, g_\gamma(\xbold)$ 
    with a factor $\xi$.
    According to Eq.~\eqref{EQ:GFunc} we write this function as
    \begin{equation}
    \xi \, g_\gamma(\xbold)
    =
    \ue^{\gamma_\xi}\refl_\xi(\xbold) - 1,
    \label{EQ:RMM_xi}
    \end{equation}
    which defines a pair
    $(r_{\xi}(\xbold), \gamma_{\xi})$ for any $\xi$
    (up to a global factor
    which keeps the product $\ue^{\gamma_\xi}\refl_\xi(\xbold)$ constant
    and will be irrelevant in the following).
    For all $\xi$ these pairs describe different 
    reflectivity functions $r_{\xi}(\xbold)$ 
    and decay rates $\gamma_\xi$,
    but relate to the same function $g_\gamma(\xbold)$
    and thus have the same possible classical densities, as seen
    from Eq.~(\ref{EQ:RMM_decay2}).

    We now assume that
    the specific density relevant for quantum mechanics,
    i.e., that describes the mean intensity 
    of resonance states with decay rate $\gamma$,
    is the same for all related pairs $(r_{\xi}(\xbold), \gamma_{\xi})$.
    This implies that it is sufficient to solve the problem for one
    particular $\xi$ and the related pair $(r_{\xi}(\xbold), \gamma_{\xi})$.
    Furthermore we assume that for $\gamma = \ginv$
    the quantum mechanically relevant density 
    is given by the classically stable density $\rhoinv(\xbold)$,
    Eq.~(\ref{EQ:RMM_inv}).
    Applying this assumption to the related pairs
    $(\refl_\xi(\xbold), \gamma_{\xi})$, it is thus sufficient
    to search for a case where
    $\gamma_{\xi}$ is the inverse decay rate 
    corresponding to the reflectivity function $\refl_\xi(\xbold)$,
    i.e.,
    $\ue^{\gamma_{\xi}} = \int 1/\refl_{\xi}(\xbold)\,\ud\xbold$, see Eq.~(\ref{EQ:RMM_inv}).
    This occurs for some specific factor $\xi = \xi^\ast$
    and leads to the density, Eq.~(\ref{EQ:RMM_inv}),
    \begin{equation}
    \rho_{\gamma}(\xbold)
    =
    \frac{1}{\ue^{\gamma_{\xi^\ast}} \refl_{\xi^\ast}(\xbold)}
    =
    \frac{1}{1 + \xi^\ast \, g_\gamma(\xbold)}
    ,
    \label{EQ:RMM_density2}
    \end{equation}
    where the second equality follows from Eq.~(\ref{EQ:RMM_xi}).
    The factor $\xi^\ast$ is uniquely determined
    from the condition on the decay of the density, Eq.~(\ref{EQ:RMM_decay2}),
    \begin{equation}
    \int \frac{g_\gamma(\xbold)}{1 + \xi^\ast \, g_\gamma(\xbold)} \, \ud \xbold
    =
    0
    .
    \label{EQ:RMM_xi2}
    \end{equation}
    Uniqueness follows from the negative derivative with respect to $\xi^\ast$
    and existence can be shown for classically allowed decay rates,
    $\min \refl(\xbold) \leq \ue^{-\gamma} \leq \max \refl(\xbold)$.
    We emphasize that 
    $\xi^\ast$ depends on $\refl(\xbold)$ and $\gamma$.

    Summarizing,
    based on our assumptions the classical density
    \begin{equation}
    \rhoxi(\xbold) = \frac{1}{1 + \xi^\ast \, g_\gamma(\xbold)}
    \label{EQ:RMM_Limit}
    \end{equation}
    describes the mean intensity of resonance states
    with decay rate $\gamma$
    in the random matrix model with escape,
    where $g_\gamma(\xbold)$ is
    defined in Eq.~(\ref{EQ:GFunc})
    and $\xi^\ast$ is determined from Eq.~(\ref{EQ:RMM_xi2}).
    The special cases,
    $\xi^\ast(\gnat) = 0$ and
    $\xi^\ast(\ginv) = 1$,
    agree with Eqs.~(\ref{EQ:RMM_nat}) and (\ref{EQ:RMM_inv}),
    respectively.
    Note that if there is a phase-space region with full escape,
    $\refl(\xbold) = 0$, then $\ginv = \infty$,
    but for $\gamma < \infty$ Eq.~\eqref{EQ:RMM_Limit} still applies.

    We numerically find that
    resonance states of the random matrix model converge
    towards the densities $\rhoxi(\xbold)$
    as $N$ increases (not shown).
    This has been tested for several reflectivity functions $\refl(\xbold)$,
    varying on phase space and including cases
    where both full and partial escape occur in different phase-space regions.

    \section{Random vector model for partial escape}
    \label{SEC:RandomVectorModel}
    
    In this section we present a random vector model for systems with
    partial escape.
    The goal is a statistical description of the complex vector $\psi$.
    This is achieved by assuming independently distributed Gaussian complex
    entries in only one distinguished basis.
    In the following we will see that the random vector model
    implies the conjecture of Sec.~\ref{SEC:Conjecture}
    for intensities with respect to any quantum state $\varphi$.

    First, let us consider the simplest setting of random matrices with escape $\MMap$,
    introduced in Sec.~\ref{SEC:NumericalResultsRandom}.
    Without loss of generality we assume that $r(\xbold)$ is a function of
    position $q$ only, such that $\Qrefl = \op{\sqrt{r}}$
    can be chosen diagonal in position basis.
    This implies, according to Eq.~\eqref{EQ:RMM_Limit}, that for all $\gamma$
    the semiclassical densities $\rhoxi(\xbold)$ also depend on $q$, only.
    Thus, quantizing the classical densities leads to a diagonal
    representation in position basis. 
    
    We propose the following random vector model for the
    distribution of coefficients of resonance states $\psi$ for some arbitrary,
    but fixed decay rate $\gamma$.
    Let $\psi_j = \braket{q_j}{\psi}$ be the coefficients in
    position basis,
    such that $|\psi\rangle = \sum_{j=1}^N \psi_j |q_j\rangle$.
    For this, the expected mean value of the intensities
    $\Iphi[q_j] = |\psi_j|^2$ is given
    by  $\rhoxi(q_j)/N$.
    This leads to the following adaption of the CUE ensemble of random states \cite{NonVor1998}:
    For any fixed decay rate $\gamma$ we define the ensemble of random vectors $\psi = (\psi_1, \dots, \psi_N)$ as
    \begin{equation}
    \prob[\gamma]{\psi} \prod_{j=1}^{N}\ud^2\psi_j
    = \prod_{j=1}^N \left(\frac{N}{\pi \rhoxi(q_j)}
    \exp\left[-\frac{N|\psi_j|^2}{\rhoxi(q_j)}\right] \right)\ud^2 \psi_j,
    \label{EQ:RWMPDist}
    \end{equation}
    where the complex coefficients $\psi_j$ are independent and identically
    distributed according to a Gaussian with variance  ${\rhoxi(q_j)}/{N}$.
    For each $\gamma$ this definition ensures that
    the expectation value $\mathbb{E}_\gamma(|\psi_j|^2)$
    is given by $\rho_{\gamma}(q_j)/N$ and that on average we have
    normalized states, i.e.,
    $\mathbb{E}_\gamma(\sum_j |\psi_j|^2)
    = \sum_j {\rhoxi(q_j)}/{N} = 1$.
    Note that such a position dependent variance also follows from the restricted random vector model describing quantum maps with a mixed phase space
    \cite{BaeNon:p}.

    In order to derive
    for some arbitrary quantum state $\varphi$
    the distribution of intensities $\Iphi$, Eq.~\eqref{EQ:IPhi},
    we consider the overlap $\nu_\varphi = \braket{\varphi}{\psi}$.
    Since Eq.~\eqref{EQ:RWMPDist}
    describes a (circularly-symmetric) complex normal distribution
    with diagonal covariance matrix
    $C_\gamma = N^{-1}\operatorname{diag}[\rhoxi(q_1), \dots, \rhoxi(q_N)]$
    it follows that $\nu_\varphi$ is also normally distributed with
    variance given by
    $C_{\gamma,\varphi} = \braket{\varphi|C_\gamma}{\varphi}
    = N^{-1}\,\sum_{j=1}^N \rhoxi(q_j)\, |\braket{q_j}{\varphi}|^2$
    \cite{Gal2008b:p}, i.e.,
    \begin{equation}
    P_\gamma(\nu)\,\ud^2\nu = \frac{1}{\pi C_{\gamma,\varphi}}
    \exp\left(-\frac{|\nu|^2}{C_{\gamma,\varphi}}\right)\ud^2\nu.
    \label{eq:distributionBargmann}
    \end{equation}
    This implies directly that $\Iphi = |\nu|^2$ is exponentially distributed
    with mean value $\meanI  = C_{\gamma, \varphi}$,
    i.e., the conjecture stated in Sec.~\ref{SEC:Conjecture}.

    In general $\refl(\xbold)$ might depend on $q$ and $p$, such that $\Qrefl$ is diagonal in some
    different basis $\{b_i\}_{i=1}^N$ with eigenvalues $w_i$. However, transforming $\MMap$ to this basis,
    $\hat{B}^\dagger \MMapcls \Qrefl \hat{B} =  \hat{V}^\text{cue}\Qrefl_\text{diag}$, implies
    a random matrix $\hat{V}^\text{cue} := \hat{B}^\dagger \MMapcls \hat{B}$ with diagonal reflection operator
    $\Qrefl_{\text{diag}} = \operatorname{diag}(w_1, \dots, w_N)$ which can be treated as before.
    The expected mean value of the intensities {$\Iphi[b_j] =  |\braket{b_j}{\psi}|^2$} for resonance states
    of $\MMap$ thus is given by $\rhoxi(b_i)/N$
    where $\rhoxi(b_i) = [{1 + \xi_\gamma(\ue^\gamma \tilde{r}_i - 1)}]^{-1}$
    and $\tilde{r}_i = |w_i|^2$.

    Finally, let us discuss how this random vector model could be extended to
    arbitrary chaotic systems with escape.
    There are three major challenges: (i) The semiclassical structure of
    resonance states in arbitrary systems is a multifractal measure without
    density. (ii) A complete semiclassical description of these measures is
    still missing. (iii) For each decay rate $\gamma$ we expect  a different basis
    for the distribution in Eq.~\eqref{EQ:RWMPDist}.
    
    For fluctuations on phase space, the first issue is overcome by the fact
    that the quantum Husimi densities are smooth functions. Thus, for fixed
    value of $h$ their expected mean value also behaves smoothly on phase
    space. Therefore, together with the second challenge,
    this reduces to the problem of obtaining
    the correct smooth density from the semiclassical multifractal measure.
    The third issue is caused by the change of the phase-space
    structure under variation of $\gamma$ \cite{ClaAltBaeKet2019}, see Fig.~\ref{FIG:ScaledHusimis}.
    Thus the mean densities $\rhoxi$ do not have a quantum
    mechanically diagonal representation in the same basis for different decay rates.
    Therefore, in contrast to the
    random model with escape,
    it would be necessary to first obtain the specific basis for each
    $\gamma$. %
    %

    
    %

\end{document}